# A Semi-Automated Technique for Internal Jugular Vein Segmentation in Ultrasound Images Using Active Contours

E. Karami, *Student Member*, *IEEE*, M. Shehata, *Member*, *IEEE*, P. McGuire, A. Smith

*Abstract*— The assessment of the blood volume is crucial for the management of many acute and chronic diseases. Recent studies have shown that circulating blood volume correlates with the cross-sectional area (CSA) of the internal jugular vein (IJV) estimated from ultrasound imagery.

In this paper, a semi-automatic segmentation algorithm is proposed using a combination of region growing and active contour techniques to provide a fast and accurate segmentation of IJV ultrasound videos. The algorithm is applied to track and segment the IJV across a range of image qualities, shapes and temporal variation. The experimental results show that the algorithm performs well compared to expert manual segmentation and outperforms several published algorithms incorporating speckle tracking.

## I. INTRODUCTION

Estimation and monitoring of circulating blood volume is important for optimal management of hyper- and hypovolemia [1-2]. Traditional methods of blood volume assessment tend to be invasive and are generally recognized as poor predictors of fluid-responsiveness [3]. In recent year, ultrasound images of the inferior vena cava (IVC) have been used to assess volume status and guide fluid administration [4]. This approach has limitations in that a skilled operator must perform repeated measurements over time and image quality is often limited by patient factors such as body habitus, bowel gas, and depth of the IVC relative to the anterior abdominal wall.

Portable ultrasound is used for localization of the internal jugular vein (IJV) for placement of the central venous catheter [5-6]. Recently, it has been shown that cross-sectional area (CSA) of the IJV is correlated with central venous pressure (CVP) and end diastolic volume. Hence, estimates of circulating blood volume can be made from ultrasound images of the IJV [7-8]. Since the IJV is a relatively superficial vascular structure in the neck, ultrasound images are generally of higher quality than those of the IV

Active Contour (AC), also called snakes, was proposed by Kass et al [9] and is a popular interactive segmentation method for 2D medical images. AC models the desired contour as a time evolving curve and the segmentation process as an optimization over time of an adequate energy functional.

E. Karami and M. Shehata are with the Faculty of Engineering and Applied Sciences, Memorial University, Canada (email: {ekarami, mshehata}@mun.ca).
P. McGuire is with C-CORE, St. John's, Newfound and Labrador, Canada (email: peter.mcguire@c-core.ca).
A. Smith is with the Faculty of Medicine, Memorial University, Canada (email: e36ajs@mun.ca).

In [10], AC technique and its combination with speckle tracking were proposed for the segmentation of the IJV images. In [10], the initial contour points for the first frame are manually selected by an experts and for the next frames, speckle tracking is used to move the control points of the AC (snake) along the motion of the edges of the IJV. This technique is efficient only if the deformation from one frame to the next one is negligible and consequently it requires a ultrasound device with a high frame rate.

In this paper, we propose an AC-based algorithm which uses region growing technique to initialize the AC. Region growing is used to grow a region around some seed pixels and obtain an initial contour for the AC algorithm. The initial seed pixel for the first frame is manually chosen by the operator and for the next frame, it is considered to be the centroid of the contour in the previous frame. Therefore, the only restriction on the proposed algorithm is that the centroid of the IJV must not have large displacement in two consecutive frames which is a very reasonable assumption.

The paper is organized as follows: Section II presents the technical details of the proposed algorithm; Section III details the experiments and results and compares to the performance in [10]; Section IV provides concluding statements and future work.

## II. PROPOSED ALGORITHM

A. Each frame of the captured video is preprocessed with a 7×7 median filter followed by a Gaussian filter to remove noises.

B. The initial seed pixel of the first frame is manually selected. For subsequent frames, the seed seed point is assumed to be the centroid of the contour from the previous frame. The centroid is calculated using first-order moments of the contour points as

$$\begin{aligned} x_c &= \sum_{x,y \in C} xI(x,y), \\ y_c &= \sum_{x,y \in C} yI(x,y). \end{aligned} \quad (1)$$

This approach is feasible if the relative motion between the ultrasound probe and the centroid of the region of interest is not significant.

C. The region is grown from the initial seed point obtained as per step B. Region growing is the simplest of the region-based image segmentation techniques [11]. The concept of region growing algorithm is to choose an initial seed pixel and grow a region based on some feature such as intensity, gradient direction, or gradient magnitude. In this paper, we consider the intensity as the region-growing feature. The region is iteratively grown by comparing all unallocated 8-connected neighboring pixels to the region such that if the difference between intensity of the candidate pixel and mean intensity of the region is less than some threshold T, then that candidate pixel is inserted into the region. This iterative process continues until there are no unchecked candidate pixels. In this paper, we set the threshold as $T = 0.05(I_{max} - I_{min})$ where $I_{max}$ and $I_{min}$ are maximum and minimum intensities of the frame, respectively.

D. A cubic spline interpolation is used to obtain $N=32$ equidistant contour points. Without this resampling, the performance of active contour algorithm will be very poor as a result of the initial contour points being too close together. The contour generated by region growing technique tends to underestimate the actual area of the IJV.

E. The AC algorithm is used to move the contour points along the edges of the IJV and provide a continuous and smooth contour. The energy function in the proposed algorithm is defined as

$$E(C) = \int_0^1 E_{image}(C)dn + \int_0^1 E_{internal}(C)dn + E_{const}(C), \quad (2)$$

where $E_{image}(C)$ is the energy term which is attracting the curve to the image edges and is defined as

$$E_{image} = -|\nabla I(x,y)|^2, \quad (3)$$

$E_{internal}(C)$ is the energy term which forces the contour to keep its shape as regular as possible and is defined as

$$E_{internal}(C) = 2|C(n)|^2 + 2|C'(n)|^2, \quad (4)$$

and $E_{const}(C)$ is the energy term related to any problem specific constraint and is defined as

$$E_{const}(C) = 2(I(x,y) - 50)A(C), \quad (5)$$

where $A(C)$ is the area of the contour which can be calculated as follows:

$$A(C) = \frac{1}{2}\left|\sum_{n=1}^{N}(x(n+1,t) - x(n-1,t))(y(n+1,t) - y(n-1,t))\right|. \quad (6)$$

The energy function of the AC is minimized as

$$\mathbf{x}_t = (\mathbf{B} + \gamma\mathbf{I})^{-1}(\gamma\mathbf{x}_{t-1} - \kappa_t \mathbf{f_x}(\mathbf{x}_{t-1}, \mathbf{y}_{t-1}) - w_c A_x(C_{t-1}))$$
$$\mathbf{y}_t = (\mathbf{B} + \gamma\mathbf{I})^{-1}(\gamma\mathbf{y}_{t-1} - \kappa_t \mathbf{f_y}(\mathbf{x}_{t-1}, \mathbf{y}_{t-1}) - w_c A_y(C_{t-1})) \quad (7)$$

where vectors $\mathbf{x}_t$ and $\mathbf{y}_t$ indicate x and y coordinates of the contour points at iteration $t$, subscripts $x$ and $y$ indicate gradient versus them, $\mathbf{B}$ is the penta-diagonal matrix defined in [9], $\gamma = 2000$, and $\kappa_t = 0.98^{-t}$.

III. EXPERIMENT AND RESULTS

The proposed algorithm was applied to track the area of two videos captured by ultrasound equipment Sonosite M-Turbo with a linear array transducer with frequency 6-15 MHz, frame rate 30 fps, and scan depth 6cm. Each video includes 450 frames corresponding to 15 seconds of data.

Experimental data was collecting using healthy subjects with different IJV shapes corresponding to different volume status. The performance of the proposed algorithm is compared with the manual segmentation and the algorithms in [10]. The project was reviewed and approved by the Health Research Ethics Authority.

*Evaluation of extraction*

The accuracy of the proposed algorithm was evaluated using DICE coefficient to signify the level of agreement between the areas extracted by the algorithm and exprt manual segmentation. DICE coefficient, *S*, is defined as

$$S = \frac{2A \cap M}{A + M}, \quad (8)$$

where $A$ and $M$ are the CSAs of the IJV estimated from the algorithm and the manual segmentation, respectively, while $A \cap M$ is corresponding intersection.

Fig. 1 shows the contour obtained through different tracking and segmentation schemes for different frame indices. The proposed algorithm (Fig. 1-second row) accurately tracks the actual contour of the IJV but consistently underestimates the area. The third row represents the result from the AC algorithm investigated in [10] which tends to overestimate the area. Note should be made that its error gradually increases over time. The last row demonstrates the speckle tracking-based AC algorithm's (presented in [10]) limited ability to follow variations in the IJV. It is suspected that this is a result of the relatively low frame rate of 30 fps[1].

Fig. 2 represents a comparison of the DICE coefficients for the proposed and previously published algorithms [10] with manual segmentation for the first subject. The DICE coefficient of the proposed algorithm is notably higher with a mean value of 0.95 compared to 0.89 of the existing algorithms. In addition, the algorithms in [10] perform well on initial frames though tend to lose track during later stages.

Fig. 3 depicts the CSA of the IJV versus frame index for the manual segmentation, the proposed algorithm, and the algorithms in [10]. The proposed algorithm continuously tracks the manual segmentation contour though it tends to slightly underestimate the area. Alternately stated, the relative variations in the CSA of the IJV estimated by the manual segmentation and the proposed algorithm demonstrate good agreement. Additional findings suggest that the contour estimated from the proposed algorithm appears smoother than the manual segmentation contour – likely introduced by the natural fluctuations of the hand of the expert.

---

[1] In [10], the frame rate is 69 fps and therefore the variations from one frame to the next one are less than our scenario.

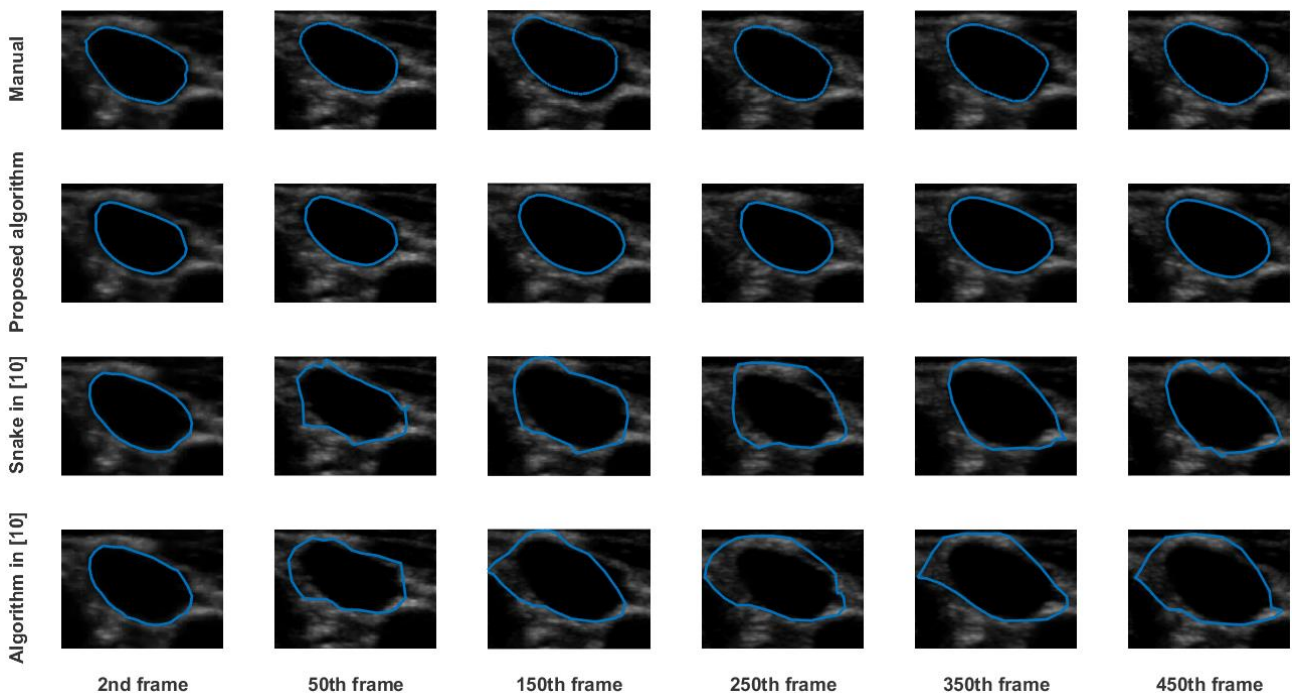

Fig. 1: Tracking of the IJV contour in different frames of the video captured from the first subject for manual segmentation, the proposed algorithm, and two algorithms proposed in [10].

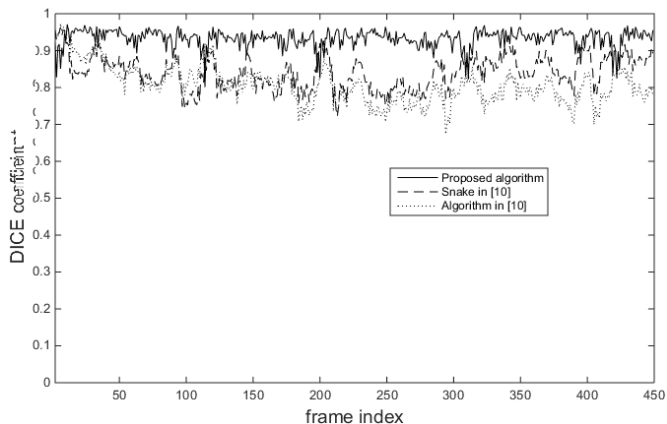

Fig. 2: The DICE factor for the proposed algorithm and the algorithms in [10] for the first subject.

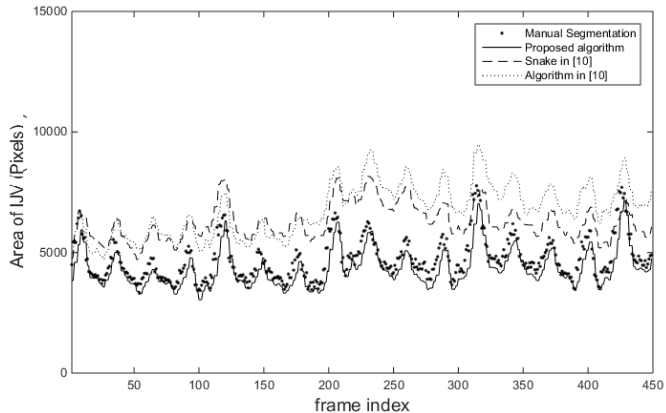

Fig. 3: The CSA of the IJV from the manual segmentation, the proposed algorithm, and the algorithms in [10] for the first subject.

Figs. 4-6 present the second subject where the IJV demonstrates near complete collapse and hence, a much lower CSA. Due to the shape of the IJV, the CSA of the IJV is more sensitive to segmentation error culminating in less accurate results, as expected. The proposed algorithm follows the variations of the IJV contour well other than areas that are fully collapsed. In comparison, the algorithms in [10] tend to lose track after the initial frames and have been shown to segment regions ouside the IJV.

For the second subject (Fig. 5), the average DICE coefficients for the proposed and [10] algorithms are 0.76 and 0.48, respectively. The [10] algorithm's performance deteriorates significantly after the first 125 frames. These findings are supported in Fig. 6. In both investigated scenarios, the proposed algorithm demonstrated superior performance.

## IV. CONCLUSION

In this paper, an active contour (AC) based segmentation algorithm was proposed for estimation and tracking of the cross-sectional area of the IJV from ultrasound images captured at 30 fps. The proposed algorithm was applied to two ultrasound videos with different IJV shapes. Experimental results show that the proposed algorithm performs very close to expert manual segmentation under lower frame-rate scenarios.

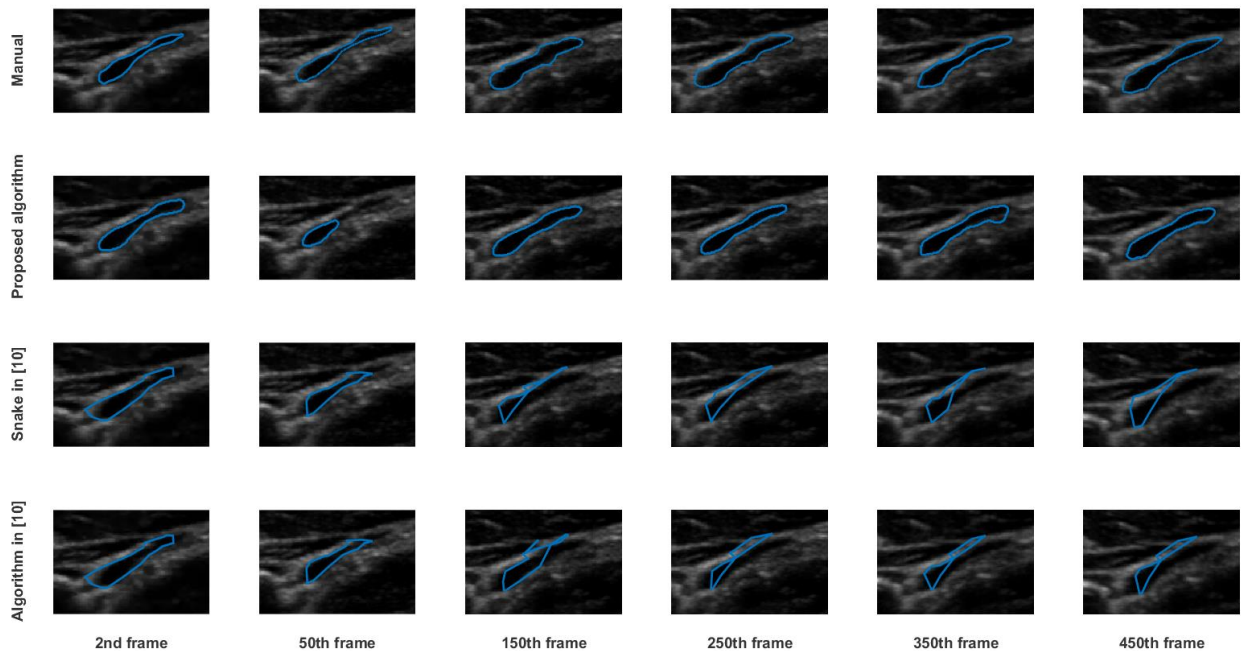

Fig. 4: Tracking of the IJV contour in different frames of the video captured from the second subject for manual segmentation, the proposed algorithm, and two algorithms proposed in [10].

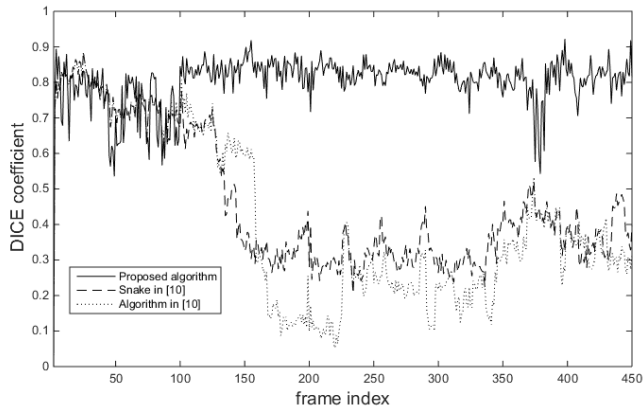

Fig. 5: The DICE factor for the proposed algorithm and the algorithms in [10] for the second subject.

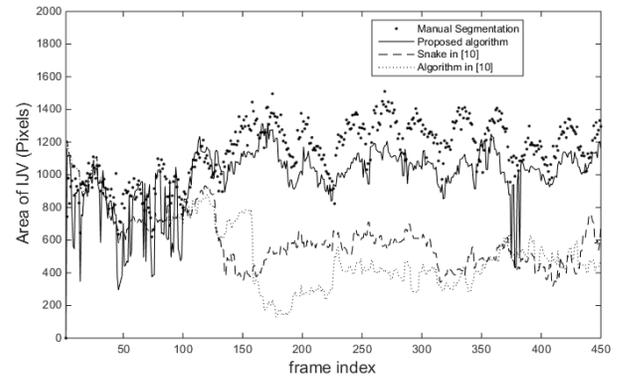

Fig. 6. The CSA of the IJV from the manual segmentation, the proposed algorithm, and the algorithms in [10] for the second subject.